\documentclass[prb,twocolumn,superscriptaddress,showpacs,preprintnumbers,amsmath,amssymb,floatfix]{revtex4}

\usepackage{epsfig}
\usepackage{graphicx}
\usepackage{color}
\usepackage{amsmath}
\usepackage[english]{babel}
\usepackage{amssymb}

\newcommand{\bra}[1]{\langle#1|}
\newcommand{\ket}[1]{|#1\rangle}
\newcommand{\beq}{\begin{equation}}
\newcommand{\eeq}{\end{equation}}
\newcommand{\Ham}{\mathcal H}
\newcommand{\sandwich}[3]{\langle#1|#2|#3\rangle}
\newcommand{\overlap}[2]{\langle#1|#2\rangle}

\begin{document}

\title{Quantum Quenches, Thermalization and Many-Body Localization}

\author{Elena Canovi}
\affiliation{International School for Advanced Studies (SISSA), Via Bonomea 265, I-34136 Trieste, Italy}

\author{Davide Rossini}
\affiliation{NEST, Scuola Normale Superiore, and Istituto Nanoscienze - CNR, Pisa, Italy}

\author{Rosario Fazio}
\affiliation{NEST, Scuola Normale Superiore, and Istituto Nanoscienze - CNR, Pisa, Italy}

\author{Giuseppe E. Santoro}
\affiliation{International School for Advanced Studies (SISSA), Via Bonomea 265, I-34136 Trieste, Italy}
\affiliation{International Centre for Theoretical Physics (ICTP), P.O. Box 586, I-34014 Trieste, Italy}
\affiliation{CNR-INFM Democritos National Simulation Center, Via Bonomea 265, I-34136 Trieste, Italy}

\author{Alessandro Silva}
\affiliation{International Centre for Theoretical Physics (ICTP), P.O. Box 586, I-34014 Trieste, Italy}

\begin{abstract}

We conjecture that thermalization following a quantum quench in a strongly correlated quantum system is 
closely connected to many-body delocalization in the space of quasi-particles. This scenario is tested in the 
anisotropic Heisenberg spin chain with different types of integrability-breaking terms. We first quantify the 
deviations from integrability by analyzing the level spacing statistics and the inverse participation ratio of 
the system's eigenstates. We then focus on thermalization, by studying  the dynamics after a sudden quench 
of the anisotropy parameter. Our numerical simulations clearly support the conjecture, as long as the integrability 
breaking term acts homogeneously on the quasiparticle space, in such a way as to induce ergodicity over all 
the relevant Hilbert space.

\end{abstract}

\pacs{75.10.Jm, 72.15.Rn, 05.45.Mt}


\maketitle

\section{Introduction}

The understanding of ergodicity and thermalization in quantum systems is one of the most 
intriguing problems in quantum physics. 
Starting with the 1929 paper of John von Neumann~\cite{vonNeumann29}, various 
attempts have been made towards the characterization of ergodic behavior in quantum
systems~\cite{Pauli37,Mazur68,Peres84,eth} and the establishment of a link 
with the notion of quantum chaos~\cite{Peres84,eth}. 
Theoretical interest in these issues resurfaced periodically~\cite{chaos_ergod}
until very recently, when an experimental 
study of the non-equilibrium dynamics of a quasi-one-dimensional condensate clearly demonstrated 
the lack of thermalization/ergodicity in a quantum many-body system~\cite{kinoshita06}. 
The attribution of this observation to quantum integrability generated 
a lot of interest on its connections with ergodicity and thermalization 
in strongly-correlated quantum systems~\cite{rigol07,quenchRefs}.

The simplest setting to study the relaxation of many-body systems is to consider an abrupt change 
in time of one of the control parameters, i.e., a {\it quantum quench}. At long times after the quench, 
the lack of thermalization in an integrable system can be seen as a consequence of the sensitivity 
to the specifics of the initial state encoded in the values of the constants of motion~\cite{rigol07}. 
This lead to the proposal of describing the time-averaged steady state reached 
after a quench by keeping track of the initial value of all the constants of motion through a 
generalized Gibbs ensemble~\cite{rigol07}, whose conditions of applicability 
and drawbacks have been extensively tested~\cite{quenchRefs}.
In turn, if the system is far enough from the integrable limit, thermalization is generally expected 
to occur, as numerically confirmed in many circumstances~\cite{rigol08,quenchNonInt,biroli}.

This qualitative picture, although very  appealing, leaves a number of important questions unanswered. 
It is not yet clear what is the nature of the integrable/non-integrable transition. 
Moreover, as it was shown in Ref.~\onlinecite{rossini09}, it appears that even 
an integrable system could look ``thermal'', depending on the observable which is analyzed. 
Operators which are non-local {\it in the quasi-particles} of the system 
may behave thermally, while local operators do not.
How to reconcile all these observations under a unifying framework?

The purpose of this paper is to show that the underlying mechanism governing 
the thermalization of many-body systems (and its relation to integrability) 
is that of {\it many-body localization} in Fock space~\cite{levitov97,basko06,gornyi05}. 
We organize the paper as follows. In Sec.~\ref{sec:MBloc}, after explaining in details 
the key concept of many-body localization, we discuss qualitatively our conjecture
on thermalization following a quantum quench, i.e., on the role played by many-body localization.
The model under investigation is then introduced in Sec.~\ref{sec:model}.
In the following sections we discuss our results: first of all we address the spectral
properties of the model and use them to characterize the localized and delocalized
regimes (Sec.~\ref{sec:spectral}); we then focus on the quench dynamics,
providing evidence of the connection between delocalization and thermalization 
according to standard statistical mechanics predictions (Sec.~\ref{sec:quench}).
Finally, in Sec.~\ref{sec:summary} we draw our conclusions.

\section{Many-body localization and quantum quenches}  \label{sec:MBloc}

In this section we discuss a qualitative scenario connecting the physics of 
thermalization after a quantum quench to the phenomenon of many-body localization. 
In order to do so, let us first lie down a few basic facts about many-body localization,
as originally discussed in the context of transport of interacting electrons in random 
potentials~\cite{levitov97,basko06}.
In absence of electron-electron interactions, the physics of disordered electron systems can be understood
in terms of the standard Anderson localization phenomenology: extended wave functions correspond
to finite zero-temperature conductivity while localized states correspond to
vanishing conductivity. The Anderson localization-delocalization transition is therefore naturally associated 
to a metal-insulator quantum phase transition. 
Notice that since localized and extended states cannot mix in the spectrum, 
the latter is a sequence of bands of extended and localized states separated by mobility edges.

The nature of the spectrum and of the eigenstates can change drastically if electron-electron interactions
are taken into account. In particular, it has been recently shown~\cite{basko06} that even when {\it all} 
single-particle states are localized, the presence of electron-electron interactions and inelastic collisions
can result in an insulator-to-metal transition as the temperature of the system is raised above a certain 
critical value $T_{c}$.
Such a phase transition can be thought of as a {\it many-body localization-delocalization transition},
occurring at the level of many-body eigenstates~\cite{levitov97,basko06}.
Indeed, the presence of the many-body localization transition at finite temperature 
implies the existence of a many-body mobility edge at an energy scaling extensively with system size and
separating localized many-body states, at low energies, from extended many-body states, 
at higher energies~\cite{basko06}. 

\begin{figure}[!t]
  \includegraphics[width=0.9\columnwidth]{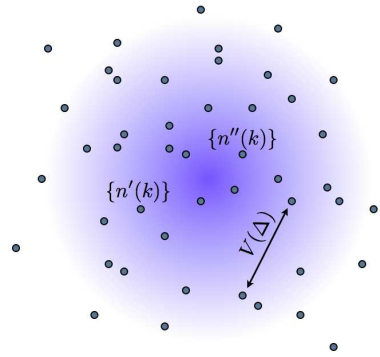}
  \caption{(color online). A cartoon of the quasi-particle space. 
    For an integrable model all states, represented by the occupations of quasi-particles $\{n(k)\}$,
    are localized. An integrability-breaking perturbation introduces hopping matrix 
    elements $V$ among different sites, which hybridize, provided $\vert E(\{n'(k)\}-E(\{n''(k)\} \vert \leq V$. 
    For strong perturbations this may lead to delocalization of wave functions among all points in  
    quasiparticle space in a microcanonical energy shell. }
  \label{fig:reticolo}
\end{figure}

Many-body localization is a rather general concept which does not necessarily refer to real space. 
A standard example in this sense comes from the physics of quasi-particle relaxation in quantum dots. 
Here the concepts of localization and delocalization find their natural applicability in Fock
space, where all many-body eigenstates are defined~\cite{levitov97}.  
Similar ideas were also employed to analyze the mixing of vibrational modes due to anharmonicity 
in molecules~\cite{logan90}. 
Along these lines, it was recently realized~\cite{pal10,prosen08,huse2} that the many-body 
localization-to-delocalization transition discussed above 
should be deeply connected to the main subject of this work, the physics of integrability-breaking. 
More specifically, let us think of an integrable model (having well defined quasi-particles) as a 
multidimensional lattice in which each point, identified by the occupations $n(k)$ of the various quasi-particle 
modes, represents an eigenstate $|\Psi_{\alpha}\rangle=|\{n_{\alpha}(k)\}\rangle$ (see Fig.~\ref{fig:reticolo}).
The space of these states (the {\it quasi-particle space}) is an obvious generalization of the 
standard Fock space.
As long as states are localized in quasi-particle space~\cite{note1}, one expects the system to behave as integrable: 
any initial condition spreads into few sites, maintaining strong memory of the initial state. On the other hand,
once a strong enough integrability-breaking perturbation hybridizing the various states 
$\ket{n_{\alpha}(k)}$ is applied, a consequent {\it delocalization in quasi-particle space} will occur
(see Fig.~\ref{fig:reticolo}). A tendency towards ergodicity is expected in this case.

In this paper we aim at establishing a close connection between the physics of the 
localization-delocalization transition/crossover, occurring in quasi-particle space 
in the presence of an integrability-breaking term, and the physics of thermalization. 
In order to do so, we focus on a specific class of non-equilibrium 
protocols on which thermalization can be studied, the so-called quantum quenches. 
In the present context they are defined through the time-dependent Hamiltonian
\beq \label{eq:model}
\Ham(t) \equiv \Ham_0[g(t)] + \Ham_{ib} \, ,
\eeq
where:
\beq \label{eq:QuenchScheme}
g(t) = \left\{ \begin{array}{ll} g_0 & {\rm for} \quad t<0 \\
                                 g   & {\rm for} \quad t \geq 0 
               \end{array} \right. \, .
\eeq
The time-dependent part of the Hamiltonian $\Ham_0[g(t)]$ is integrable, while $\Ham_{ib}$ is 
the integrability-breaking term. We then ask ourselves the following question: what are the characteristics 
that many-body eigenstates should have in order for the system to thermalize and behave ergodically?

The answer to an analogous question for semiclassical quantum chaotic systems was conjectured by M. Berry in 
1977~\cite{Berry} and later employed by M. Srednicki to discuss thermalization in a (non-integrable) gas of 
interacting particles~\cite{eth}. 
Inspired by these seminal papers, we propose that for generic many-body systems thermalization will occur 
whenever the eigenstates of the system become {\it diffusive} in microcanonical shells defined in quasi-particle space. 
These diffusive states correspond to the intuitive expectation that in an ergodic state 
any initial state is allowed to diffuse into all states in a micro-canonical energy shell, generating a cascade 
of all possible lower energy excitations~\cite{note2}. 
The purpose of the remaining sections is to test this proposal on the dynamics of a concrete integrable model.

\section{The Models}   \label{sec:model}

In order to corroborate the scenario proposed above, we now study in detail the dynamics after a 
quench of an anisotropic Heisenberg chain subject to various forms of integrability-breaking perturbations.
The anisotropic Heisenberg (XXZ) spin-$1/2$ chain is defined by:
\beq \label{eq:XXZ}
\Ham_0(J_z) = \sum_{i=1}^{L-1} \Big[
   J \left(\sigma^{x}_{i}\sigma^{x}_{i+1}+\sigma^{y}_{i}\sigma^{y}_{i+1}\right)
    + J_{z}\sigma^{z}_{i}\sigma^{z}_{i+1} \Big] \,,
\eeq
where $\sigma^\alpha_i$ ($\alpha=x,y,z$) denote the spin-$1/2$ Pauli matrices on site $i$,
$J$ the coupling strength, while $J_z$ the $z$-anisotropy intensity. 
Units of $\hbar = k_B = 1$ are used throughout the paper; $J = 1$ is taken as the energy scale.
This Hamiltonian is integrable by Bethe Ansatz, and exhibits two gapped phases, 
ferromagnetic ($J_{z}<-1$) and antiferromagnetic ($J_{z}>1$), separated 
by a critical region $-1\leq J_{z}\leq 1$, with $J_{z}$-dependent critical 
exponents~\cite{Haldane_PRL81} and quasi-long-range-order in the $xy$ spin-plane. 

As for the integrability-breaking perturbation, we consider different cases,
with or without disorder terms in the Hamiltonian,
which can be expressed in the form:
\beq
   \Ham_{ib}=  \sum_i \Delta_i {\cal O}_i \, ,
\eeq
where $\Delta_i$ is the amplitude (possibly site-dependent) of an additional few-body term ${\cal O}_i$. 
This few-body term may act on a single site $i$ (e.g., onsite magnetic field), or on a few sites centered 
around $i$ (e.g., nearest or next-to-nearest neighbor couplings).
In particular, we break integrability by either adding: $(I)$ a random magnetic field in the $z$-direction; 
$(II)$ random $J_z$ couplings; $(III)$ random or $(IV)$ uniform next-nearest neighbor $zz$ couplings,
according to:
\beq
\nonumber
\Ham_{ib} = \left\{ \begin{array}{lc}
\Delta \sum_{i=1}^{L}h_{i}\sigma^{z}_{i}                    \quad & (I)   \vspace*{2mm} \\
\Delta \sum_{i=1}^{L-1}h_{i}\sigma^{z}_{i}\sigma^{z}_{i+1}  \quad & (II)  \vspace*{2mm} \\
\Delta \sum_{i=1}^{L-2}h_{i}\sigma^{z}_{i}\sigma^{z}_{i+2}  \quad & (III) \vspace*{2mm} \\
\Delta \sum_{i=1}^{L-2}\sigma^{z}_{i}\sigma^{z}_{i+2}       \quad & (IV)
\end{array}
\right. \; .
\eeq
For cases $(I)$-$(II)$-$(III)$ integrability-breaking is induced by the disorder, 
$h_i \in [-1,1]$ being random numbers, while in case $(IV)$ disorder is not invoked.
In general, disordered systems allow for a better statistical analysis, due to the possibility
of averaging over randomness.
One might argue that such averages are strictly required in order to reproduce
our findings about thermalization. 
This is not the case, since we have found analogous qualitative conclusions in all the four cases
discussed above: integrability-breaking is the only crucial requirement 
for our mechanism of thermalization to set in.

We first address the spectral properties of the model,
and subsequently consider a sudden quench of the anisotropy parameter $J_{z} \equiv g$.
The total magnetization $S^{z}=\sum_{i} \sigma^{z}_{i}$ is a conserved quantity,
hence we restrict to the sector $S^{z}=0$. 
Nonetheless, due to the involvement in such non-equilibrium dynamics of a considerable part of the spectrum, 
standard (both analytic and numerical) renormalization group techniques are eventually doomed to failure.
We therefore resort to exact numerical diagonalization of systems with up to 16 spins.

While the zero-temperature phase-diagram in presence of disorder is well established~[\onlinecite{Doty92}], 
the high-temperature phase-diagram has been conjectured to be composed of two phases, 
a non-ergodic many-body localized phase (in real space) at $\Delta > \Delta^{\rm crit}$, 
and an ergodic one at $\Delta < \Delta^{\rm crit}$; 
in the case $(I)$, $\Delta^{\rm crit} \sim 6\div 8$ at $J_z=1$ (in our units)~\cite{pal10,prosen08}. 
The results presented below indicate the presence of a second non-ergodic localized phase (in quasi-particle space) 
for $\Delta$ close to zero that crosses over to the ergodic phase upon increasing $\Delta$. 
The fate of this crossover in the thermodynamic limit and the eventual value of the critical $\Delta^*$ 
are yet to be determined~\cite{note3}.  

We start by characterizing deviations from integrability in 
terms of the many-body level statistics and of the properties of the eigenstates.
A well defined transition from Poisson (Integrable) to Wigner-Dyson statistics (non-Integrable) 
is closely associated to the localized/diffusive character of eigenstates in quasi-particle space.
Using this characterization, we then show that the non-thermal-to-thermal transition
in the dynamics is directly connected to the localization/delocalization transition 
in quasi-particle space. In particular, by looking at the asymptotics of spin-spin correlation functions, 
we discuss how thermalization is linked to the emergence of diffusive eigenstates in quasi-particle space. 
This also allows us to discuss, in a broad context, the relationship between locality of observables 
in quasi-particle space and the corresponding behavior.

\section{Spectral properties} \label{sec:spectral}

Let us first concentrate on the spectral properties of the Hamiltonian
for a given value of the anisotropy $g = J_z$. In the following we show data for $J_z = 0.5$. 
We have explicitly checked that, changing $J_z$ to a different value within
the critical region of the XXZ model ($\vert J_z \vert \leq 1$), does not qualitatively
affect the scenario discussed and our conclusions.

\subsection{Level Spacing Statistics}

The statistics of the energy levels represents a key feature of the spectrum of 
a generic quantum system, since it is a good indicator of the presence of integrability.
Both in semiclassical and in many-body systems, integrable systems 
have levels that tend to cluster, eventually crossing when a parameter in the Hamiltonian
is varied. On the other hand, in non-integrable systems the levels are
correlated in such a way as to avoid crossings.
A quantitative way to characterize these tendencies is through the 
{\it Level Spacing Statistics} (LSS)~\cite{Haake}, 
i.e., the probability distribution $P(s)$ that the energy 
difference between two adjacent levels $s_n \equiv E_{n+1}-E_n$ (normalized 
to the average level spacing) falls in the interval $[s,s+ds]$.
In a typical integrable system one finds a Poissonian (P) LSS:
\beq \label{eq:poisson}
P_{\rm P} (s) = e^{-s} \;.
\eeq
On the other hand, for non-integrable systems one expects random matrix theory to apply,
leading to a Wigner-Dyson (WD) distribution, where level repulsion shows up in
$\lim_{s \to 0} P(s) \sim s^{\gamma}$.
More specifically, for systems as the one considered here,
which preserve one anti-unitary symmetry (invariance under time-reversal),
the statistics is given by a Gaussian Orthogonal Ensemble~\cite{Haake} 
(at low energy spacings one has the characteristic behavior $\gamma = 1$):
\beq \label{eq:WD}
P_{\rm WD} (s) = \frac{\pi s}{2} \, e^{-\frac{\pi s^2}{4}} \,.
\eeq

In our case the system undergoes a transition from Poissonian [Eq.~\eqref{eq:poisson}] to Wigner-Dyson 
[Eq.~\eqref{eq:WD}] LSS upon increasing the non-integrable perturbation $\Delta$, which for finite-size systems 
takes the form of a smooth crossover. 
This can be faithfully quantified by means of the Level Spacing Indicator (LSI) $\eta$:
\beq \label{eq:eta}
\eta\equiv\frac{\int_{0}^{s_0}[P(s)-P_{P}(s)]ds}{\int_{0}^{s_0}[P_{WD}(s)-P_{P}(s)]ds} \,,
\eeq
where $P(s)$ is the probability distribution function of the level spacing between neighboring levels,
while $s_{0}\approx 0.4729$ is the first intersection point of $P_{\rm P}(s)$ and $P_{\rm WD}(s)$. 
The LSI is zero for systems with a Poisson distribution $P_{\rm P}$ of the spacings and one
if the distribution is Wigner-Dyson $P_{\rm WD}$. 
Below we will study the LSI in two different ways: 
$(i)$ as a function of the energy eigenvalues in individual microcanonical shells [$\eta_{\rm w} (E)$], 
using Eq.~\eqref{eq:eta} with $P_{[E,E+W]}(s)$ as the level statistics computed in the energy window $[E,E+W]$;
$(ii)$ cumulatively [$\eta_{\rm c} (E)$], as a function of the energy eigenvalues below a given threshold,
with $P_{[E_0,E]}(s)$ as the level statistics of eigenvalues with excitation energy less than $E$, 
with respect to the ground state energy $E_0$.

\subsubsection{Results for the LSI}

The LSS in the XXZ model with an integrability-breaking perturbation
has been the subject of various studies in the literature~\cite{Distasio95,lssXXZ}. 
Here we are not interested in a complete characterization of it, but rather on 
elucidating under which conditions, and in which regions of the energy spectrum, 
the model behaves according to the WD statistics, i.e., $\eta \approx 1$
according to our definition.
We point out that the WD distribution of Eq.~\eqref{eq:WD} is obtained for non-integrable
systems with only a time-reversal symmetry.
In all our simulations we considered open boundary conditions, fixed the sector of
zero magnetization, and added a very small magnetic field on the first site of the chain, 
in such a way as to work in a subspace without any other unwanted symmetry.

\begin{figure}[!t]
  \begin{center}
    \includegraphics[width=1.\columnwidth]{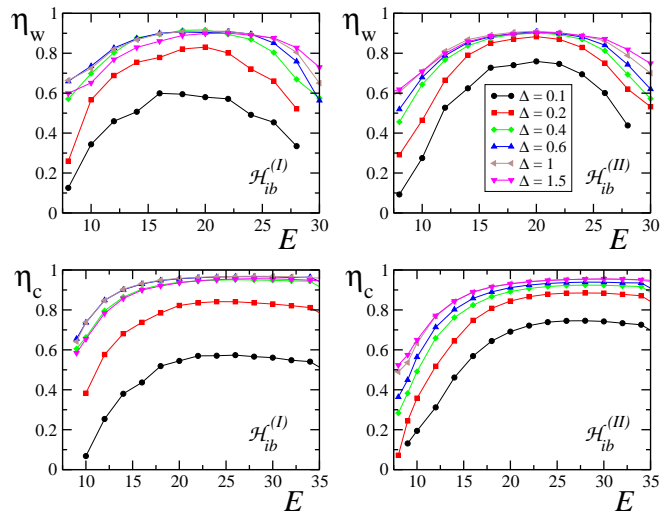}
    \caption{(color online). Level spacing indicator for the XXZ model~\eqref{eq:XXZ} with 
      non-integrable perturbations $(I)$ (left panels) and $(II)$ (right panels).
      The upper frames show $\eta_{\rm w} (E)$, while the lower ones display $\eta_{\rm c} (E)$.
      Data are for $L = 14$ sites with different values of the integrability-breaking perturbation $\Delta$.
      The LSI $\eta_{\rm w} (E)$ is evaluated in a microcanonical shell of width $W = 2$.
      Averages are performed over $10^3$ disorder instances.
      In order to exactly recover Poisson and GOE statistics in the two integrable and
      non-integrable limits, we performed an unfolding of the energy spectrum for each instance, 
      according to standard techniques adopted in quantum chaos~\cite{Haake}.}
    \label{fig:Eta}
  \end{center}
\end{figure}

All the four types of integrability-breaking perturbation behave quite in the same way,
the only difference being for case $(IV)$, where fluctuations
are more consistent, due to the absence of disorder averaging
(to reduce fluctuations, one should consider energy spectra of larger systems;
however, the exact diagonalization technique intrinsically imposes severe size limitations).
As an explicative example, in Fig.~\ref{fig:Eta} we plot both $\eta_{\rm w}(E)$ and
$\eta_{\rm c}(E)$ for the XXZ model with a random $z$-field, $(I)$
(left panels), and with random $J_z$ couplings, $(II)$ (right panels).
We observe that, fixing the system size, if $\Delta$ is progressively increased, 
the value of $\eta$ also increases, until it reaches, in the middle of the energy band,
a value close to $1$ (for $\Delta \sim 1$, in our units and at $L=14$).
For $\Delta \gtrsim 1$, $\eta$ decreases again towards small values, since 
for $\Delta \gg J_z$ the system turns into a trivial classically integrable model~\cite{lssXXZ}.
Only in the middle of the spectrum the system appears to exhibit
level repulsion, while this is not the case in the low- or high-energy spectrum~\cite{Distasio95}.
This is more evident from the cumulative LSI $\eta_c$;
here one can notice that, for sufficiently strong perturbations and at low energies, 
$\eta_c$ is an increasing function of $E$, until it saturates around its maximal value.

\subsection{Inverse participation ratio}

After a spectral characterization of integrability through the LSS,
we come back to the characterization of the eigenstates. 
In the region where a Poissonian LSS is observed, the eigenstates are expected to be localized
in quasiparticles space, while if WD is seen, the eigenstates are expected to be delocalized. 
The proper tool to quantitatively characterize the properties of the eigenstates and their 
delocalization is the so called {\it Inverse Participation Ratio} (IPR)~\cite{Haake,chaosXXZ}. 
The IPR on a normalized pure state $\ket{\psi}$ is a basis-dependent quantity, defined by:
\beq \label{eq:ipr}
\xi(\ket{\psi}) = \frac{1}{N} \left(\sum_{n=1}^{N}|\overlap{n}{\psi}|^{4}\right)^{-1} \,,
\eeq
where $\lbrace\ket{n}\rbrace$ is the reference basis of the Hilbert space.
If a state is a uniform superposition of $n_{\rm st}$ basis states, the corresponding
contribution to $\xi$ is of order $n_{\rm st}$. 

We will focus on the IPR of the system eigenstates, evaluated of two types of basis: 
$(i)$ the site $(S)$ basis $\ket{n_S}=\ket{\sigma_1\cdots\sigma_L}$ ($\sigma_i=\pm1$), 
composed by the eigenstates of $\sigma^z_i$, which is often referred to
as the ``computational basis''; 
$(ii)$ the integrable $(I)$ basis, composed by the eigenstates of the integrable
model~\eqref{eq:XXZ} in absence of the perturbation terms: $\Delta=0$. 
Analogously to the LSI, we can compute the IPR over microcanonical shells around a given
energy value $E$. Notice that, if an eigenstate is localized in quasi-particle space, 
we expect the inverse participation ratio computed in the integrable basis to be $\xi_I \simeq O(1)$.
Conversely, if an eigenstate is a diffusive superposition with random phases and similar amplitudes 
of $N$ eigenstates of the integrable model, then $\xi_I \simeq N$. 
Below we will use these facts to characterize localization and delocalization in quasi-particle space.

\subsubsection{Results for the IPR}

\begin{figure}[!t]
  \begin{center}
    \includegraphics[width=1.\columnwidth]{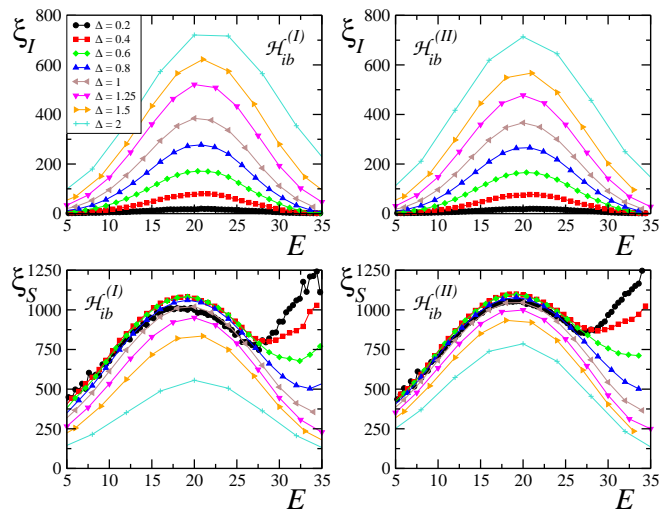}
    \caption{(color online). Inverse participation ratio for the XXZ model~\eqref{eq:XXZ} 
      with non-integrable perturbations $(I)$ (left panels) and $(II)$ (right panels).
      Data are for $L = 14$ sites and different values of the perturbation strength $\Delta$.
      The IPR is evaluated in a microcanonical shell of width $W = 2 \Delta$, 
      in the integrable (upper frames) and in the site (lower frames) basis. 
      Data are averaged over all the system eigenstates in the appropriate energy window, 
      and over $10^2$ disorder realizations.}
    \label{fig:ipr}
  \end{center}
\end{figure}

In Fig.~\ref{fig:ipr} we show the inverse participation ratio for the XXZ model 
with the integrability-breaking terms $(I)$ and $(II)$ (as for the LSI, results 
are not qualitatively different if different perturbations are considered).
Looking at the IPR in the site basis (lower panels), as long as $\Delta$ is increased
we observe a general tendency to a localization (the IPR peak value decreases). 
This is coherent with the fact that the states 
of the computational basis are exactly the eigenstates of the system for $\Delta \gg J_{z}$.
On the other hand, as depicted in the upper panels, the IPR in the integrable basis 
behaves rather differently.
In particular, it provides a clear signature of the fact that eigenstates delocalize 
with increasing values of the disorder $\Delta$. 

\begin{figure}[!t]
  \includegraphics[width=1.\columnwidth]{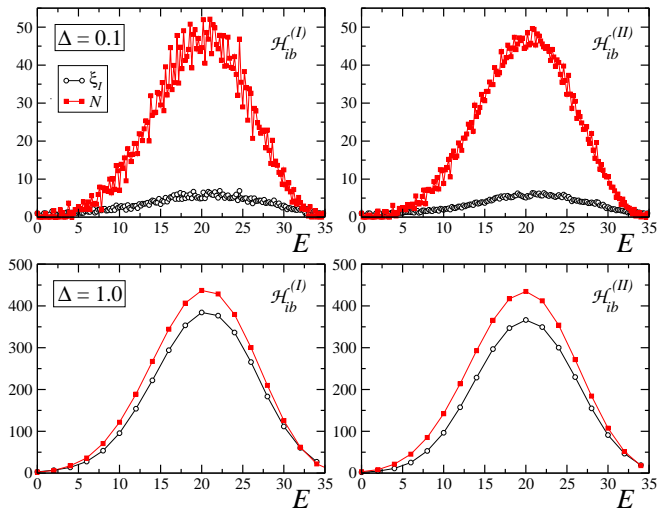}
  \caption{(color online). IPR in the integrable basis $\xi_I$ at $\Delta = 0.1$ (upper panels), 
    and at $\Delta = 1$ (lower panels), compared to the number of states $N$ 
    in an energy window of width $W = 2 \Delta$.
    Results are shown for the integrability-breaking terms $(I)$ (left panels)
    and $(II)$ (right panels). 
    Notice however that all other cases display an identical qualitative behavior.
    In particular, the disorder is not required to observe 
    the hybridization of the quasiparticle states within 
    the microcanonical energy shell for the case of large $\Delta$.}
  \label{fig:xi_N}
\end{figure}

In order to better understand the nature of the delocalization induced by the integrability-breaking term, 
let us compare $\xi$ with the number of states $N_{[E,E+W]}$ in the relevant microcanonical shell $[E,E+W]$, 
where $W\sim V$ is the typical matrix element of the integrability-breaking perturbation
($V \approx 2 \Delta$ in this case). 
For small $\Delta$, $\xi_{I}\ll N_{[E,E+W]}$ (see Fig.~\ref{fig:xi_N}, upper panels), indicating 
that the eigenstates are still close to those of the integrable system
and the degree of delocalization of the system is very low.
On the contrary, when $\Delta \simeq 1$ (Fig.~\ref{fig:xi_N}, lower panels) we observe that 
$\xi_{I} \simeq N_{[E,E+W]}$. In this case, the perturbation is able to hybridize 
nearly all the quasiparticles states within the microcanonical energy shell.
As we will see in the next section, this is the key ingredient for the system to thermalize.
Notice that in this context the low-lying eigenstates are rather peculiar: this part of the spectrum, 
which contains very few states as compared to the center, 
has closely Poissonian statistics and is characterized by large fluctuations of statistical quantities.

\section{Dynamics after the quench} \label{sec:quench}

It is now time to use the information we obtained about spectral statistics and 
eigenstates to study  the relation between many-body localization and thermalization. 
We will look at the dynamics following a sudden quench of the anisotropy parameter $g \equiv J_z$ 
from $J_{z0}$ at $t\leq 0$ to $J_{z} \neq J_{z0}$ at $t>0$,
as described in Eq.~\eqref{eq:QuenchScheme}. 
As we did for the spectral properties, we will show data for systems
where the anisotropy is quenched toward $J_z = 0.5$.
Different values of such $J_z$ do not qualitatively affect the scenario.
The system is initially prepared in the ground state $\ket{\psi_{0}}$ of $\Ham(J_{z0})$, 
so that its (conserved) energy with respect to the final Hamiltonian $\Ham(J_z)$ is 
$E_{0} = \bra{\psi_{0}}\Ham(J_{z})\ket{\psi_{0}}$.
For growing values of $J_{z0}$, the state $\ket{\psi_0}$ tends towards the classical antiferromagnetic
N\'eel state, and $E_0/L$ saturates to a constant value, slightly below the middle of the spectral band, thus
implying that a quench generally involves only a fraction of the eigenstates of the final Hamiltonian.

\subsection{Effective temperature}

Contrary to local quenches, the work done on the system by changing the anisotropy from
$J_{z0}$ to $J_z$ is extensive. 
It is then interesting to ask, after a quench involving an extensive injection of energy $E_{0}-E_{\rm gs}\propto L$
[$E_{\rm gs}$ being the ground state energy of $\Ham(J_z)$], 
if the subsequent long-time evolution of the system is effectively described by an {\em equilibrium} dynamics
governed by $\Ham(J_z)$. 
In view of a plausible equivalence between a microcanonical (fixed $E_0$) and a canonical equilibrium 
description of such a long-time dynamics, it is meaningful to define, 
as in previous instances~\cite{rossini09}, an {\it effective temperature} $T_{\rm eff}$ 
for the system out of equilibrium.
We compute $T_{\rm eff}$ by equating the micro-canonical energy 
$E_0 = \langle \psi_0 \vert \Ham (J_z) \vert \psi_0 \rangle$ to the canonical ensemble average
\beq \label{eq:teff}
E_{0} \equiv \langle\Ham(J_z)\rangle_{T_{{\rm eff}}} = {\rm Tr} \left[ \rho(T_{\rm eff}) \, \Ham(J_z) \right] \, ,
\eeq
where $\rho(T_{\rm eff})$ is the equilibrium density matrix at temperature $T_{\rm eff}$:
\beq
 \rho(T_{\rm eff}) = \frac{ e^{- \Ham(J_z) / T_{\rm eff}} }{{\rm Tr} [ e^{- \Ham(J_z) / T_{\rm eff}} ]} \, .
\eeq  
This temperature is eventually averaged over disorder realizations in the cases $(I)$-$(II)$-$(III)$.

\begin{figure}[!t]
  \begin{center}
    \includegraphics[width=1.\columnwidth,height=0.75\columnwidth]{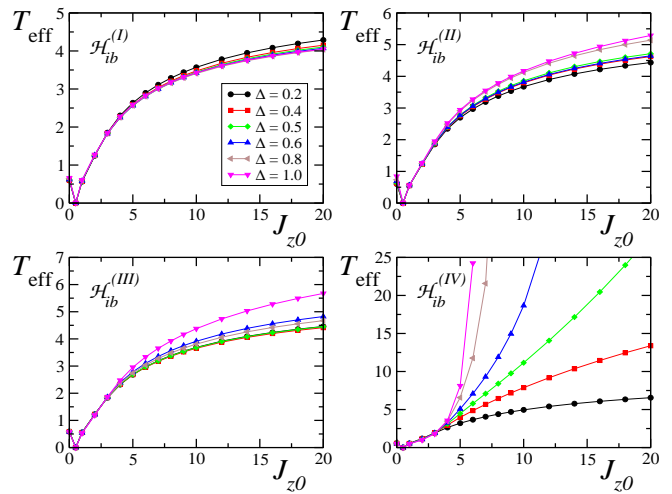}
    \caption{(color online). Effective temperature in the XXZ model with 
      an integrability-breaking perturbation,
      after a quench in the anisotropy parameter toward a value of $J_z = 0.5$.
      The values of $J_{z0}$ in the $x$-axis denote the initial anisotropies, and
      stand for different initial conditions [that is, the ground states of the
      Hamiltonian $\Ham(J_{z0})$]. 
      Data are for $L = 12$ sites; in all panels, except the lower right one,
      averages are performed over 200 disorder instances.}
    \label{fig:effTxxz}
  \end{center}
\end{figure}

In Fig.~\ref{fig:effTxxz} we show the effective temperature as a function of the initial value 
of the anisotropy for a system of $L=12$ sites, quenched toward $J_z = 0.5$. 
As it is apparent, $T_{\rm eff}$ is monotonically increasing with $\vert J_z - J_{z0} \vert$. 
In the first three cases, the effective temperature saturates for large values of $J_{z0}$, because 
the initial ground state $\vert \psi_0 \rangle$ tends toward the antiferromagnetic N\'eel state 
(for $J_{z} \gg 1$ and $\Delta \lesssim 1$
the effective temperature is around $T_{\rm eff}\sim 5$, thus meaning that
the states probed are located in the lower central part of the band)~\cite{not3}.

\subsection{Thermalization of correlation functions}

We are now ready to test the relation between delocalization and thermalization after a quench 
by studying the long-time asymptotics of two-spin correlators, constructed as expectation values of 
\beq
n^{\alpha}_k \equiv \frac{1}{L} \sum_{j,l=1}^{L} e^{2\pi i (j-l)k/L}
  \sigma^{\alpha}_{j} \sigma^{\alpha}_{l} \, , \qquad (\alpha = x,z).
\eeq
In particular, we compare the expectation value in the canonical ensemble 
at the corresponding $T_{\rm eff}$:
\beq
n^{\alpha}_{T_{\rm eff}}(k) \equiv \langle n^{\alpha}_k \rangle_{T_{\rm eff}}
= {\rm Tr} \left[ \rho(T_{\rm eff}) \, n^{\alpha}_k \right] \,,
\eeq 
with the asymptotic value that is reached after the quench, calculated from the 
diagonal ensemble~\cite{rigol08}: 
\beq
n_{Q}^{\alpha}(k) \equiv \lim_{t\to \infty} \bra{\psi(t)} n^{\alpha}_k \ket{\psi(t)}
   = \sum_{i}|c_{i}|^{2}\sandwich{\phi_{i}}{n^{\alpha}_k}{\phi_{i}} \,,
\eeq
where $\ket{\psi(t)} = e^{-i \Ham(J_z) t} \ket{\psi_0}$ is the state of the system at time $t$,
while $c_{i} = \langle \phi_{i}|\psi_{0} \rangle$ is the scalar product between the state
$\ket{\psi_0}$ and the eigenstates $\ket{\phi_i}$ of the final Hamiltonian $\Ham(J_z)$.

\begin{figure}[!t]
  \begin{center}
    \includegraphics[width=1.\columnwidth]{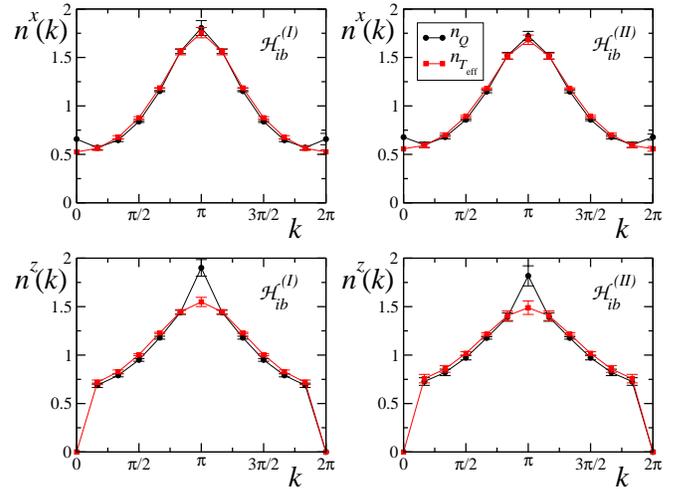}
    \caption{(color online). Comparison between the diagonal and canonical expectation value 
      of the two-spin correlation function $n^x(k)$ (upper panels)
      and $n^z(k)$ (lower panels) as a function of the momentum $k$. 
      Data are for a quench from $J_{z0} = 10$ to $J_z = 0.5$ and two kinds of
      integrability-breaking perturbation [$(I)$ in left panels, and  $(II)$ in
      right panels], with disorder intensity $\Delta=0.4$.}
    \label{fig:nx_nz}
  \end{center}  
\end{figure}

The observables we consider here correspond to two completely different scenarios
in terms of the system quasiparticles, being $n^x_k$ a {\it local} operator
while $n^z_k$ a {\it non-local} one.
We recall that, in this context, local and non-local operators refer to the 
structure of their matrix elements on the basis of quasi-particles: local means 
that the operator couples a finite number of states, while non-local 
that it couples all states~\cite{rossini09}. 
While correlators in the $x$-direction are always well reproduced by an effective thermal ensemble, 
correlators in the $z$-direction appear to be more sensitive to the breaking of integrability.
This is seen quite clearly in Fig.~\ref{fig:nx_nz}, where we plot the correlators $n^\alpha_k$
averaged in the diagonal (black circles) and in the canonical (red squares) ensembles, 
both along the $x$-axis (upper panels) and along the $z$-axis (lower panels). 
The parameters are chosen in such a way as to have the system close to integrability, 
with a significant delocalization in Fock space still not present.
Two different behaviors for $n^{x}(k)$ and $n^{z}(k)$ are apparent, with differences clearly 
emerging at the peaks $k=\pi$, where boundary effects are less pronounced.
One can qualitatively see that, while discrepancies between the two ensembles
are well visible in $n^{z}(k)$, they are suppressed in $n^{x}(k)$.
Therefore, in a quasi-integrable regime only $n^x$ behaves thermally, while $n^z$ does not.
As stated above, this reflects the intrinsic difference between nonlocal/local 
operators with respect to the quasiparticles, which emerges only for the cases
in which the system itself is not able to properly hybridize the quasiparticle states within
the microcanonical shell. 
Here we stress however that the classification of the operators is in general a subtle issue. 
In the model we considered in this work, it has been possible by analyzing the XX limit ($J_z = 0$) 
and the low-energy sector of the critical phase~\cite{Nagaosa}.

\begin{figure}[!t]
  \begin{center}
    \includegraphics[width=1.\columnwidth]{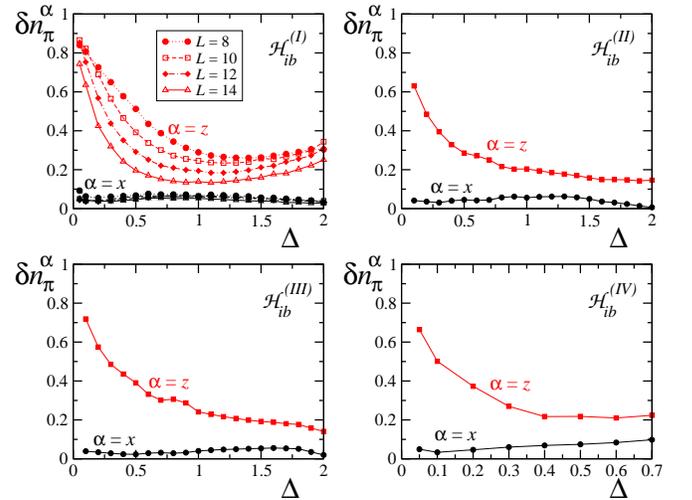}
    \caption{(color online). Discrepancies $\delta n^{x}_\pi$ (black circles) 
      and $\delta n^{z}_\pi$ (red squares) between the diagonal and the canonical
      ensemble predictions, for a quench  from $J_{z0} = 10$ to $J_{z}=0.5$. 
      Data in the upper left panel $(I)$ are for different system sizes as depicted 
      in the caption, while in the other panels are for $L = 12$.
      In the cases of perturbations involving disorder, $(I)$-$(II)$-$(III)$, 
      averages over 200 instances are performed.}
    \label{fig:deltaN}
  \end{center}  
\end{figure}

A quantitative measure of the degree of thermalization is given by the absolute discrepancy
between the diagonal and the canonical ensemble predictions:
\beq
\delta n^{\alpha}_k = |n^{\alpha}_{Q}(k)-n^{\alpha}_{T_{\rm eff}}(k)| \,.
\eeq
In order to elucidate the drastically different behavior between integrable and non-integrable systems, 
in Fig.~\ref{fig:deltaN} we plot $\delta n^{\alpha}_k$ at the peak $k=\pi$ where discrepancies are larger, 
as a function of the disorder amplitude $\Delta$.
%
%
We observe that ${\delta n}^x_{\pi}$ is more than one order of magnitude smaller 
than ${\delta n}^z_{\pi}$, indicating a closely thermal behavior, 
while ${\delta n}^z_{\pi}$ shows a sharp decrease as integrability is progressively broken 
by increasing $\Delta$, towards a minimum value at $\bar{\Delta}$.
The scaling with the dimension $L$ of the chain, shown in the upper left panel of Fig.~(\ref{fig:deltaN}), 
confirms our predictions. While the behavior of ${\delta n}^x_{\pi}$ 
is independent on the system size, the decrease  of ${\delta n}^z_{\pi}$ as a function of $\Delta$ 
is more pronounced on increasing $L$.
Due to the numerical limitations of exact diagonalization,
we cannot rule out the possibility that, in the thermodynamic limit, the integrable to non-integrable
transition for low perturbation strengths occurs at $\Delta^* = 0$ in all the cases analyzed here $(I)$-$(IV)$. 
However, for the considered sizes, we found that the local observable ${\delta n}^z_{\pi}$ thermalizes
the best at around $\bar{\Delta} \sim 1$ for model $(I)$. 
It is now crucial to observe that around this point, as noticed in the previous sections,
the diffusive nature of the eigenstates is pronounced. 
The upper right and the lower left panels seem to locate the optimal thermalization point 
for models $(II)$ and $(III)$ at a slightly larger value of $\Delta$,
while the lower right panel shifts it to slightly smaller values for model $(IV)$.
In all cases it is however true that at these points $\xi  \simeq N_{[E,E+W]}$, making a direct connection
between good thermalization and diffusive nature of the eigenstates.
Notice also that, for $\Delta \gtrsim \bar{\Delta}$, ${\delta n}^z_{\pi}$ necessarily has to increase again, 
in agreement with the fact that $\Delta \simeq \bar{\Delta}$ is the point where the non-integrable behavior 
is most pronounced and that for large values of $\Delta$ the system tends toward another integrable limit~\cite{note4}.  
In analogy with previous studies, the different sensitivity to integrability of 
correlators in different spin directions can be qualitatively understood as a consequence of the fact
that $\sigma^z$ is a local operator in quasi-particle space while $\sigma^x$ is a non-local one~\cite{rossini09}.

\section{Summary} \label{sec:summary}

In conclusion, we discussed thermalization and integrability-breaking in the dynamics after a quench 
of a quantum XXZ Heisenberg spin chain in presence of an integrability-breaking term.
We have shown that, if one wants to know when and how an interacting many-body system thermalizes, 
one should study the corresponding many-body localization/delocalization transition in 
quasi-particle space. Thermalization should occur when the relevant typical states spread
diffusively on an exponential number of states lying in the microcanonical energy shell. 
We point out that our picture is valid as long as the integrability-breaking term acts 
homogeneously in the quasiparticle space, in such a way as to induce ergodicity over all the relevant 
Hilbert space. For generic dynamic systems there may be regions of the phase space
which are non chaotic, so that their quantum versions produce entropy at a non-uniform rate given by
the local Lyapunov exponents~\cite{monteoliva}. In this case more complex scenarios for the approach 
to equilibrium may arise.

\acknowledgments

We thank B. Altshuler, V. Kravtsov and D. Huse for fruitful discussions 
and useful comments on the manuscript. 
We also benefited from discussions with T. Caneva, M. M\"{u}ller, G. Mussardo, 
A. Nersesyan, T. Prosen, A. Scardicchio, and M. \v Znidari\v c.



\begin{thebibliography}{100}
\bibitem{note1}
  Localization is intended in the space of quasi-particles, and not necessarily in real space.
\bibitem{note2} 
  Notice that in general it is possible to have non-ergodic, but extended states 
  (see Ref.~\onlinecite{levitov97} and Refs. therein).
\bibitem{note3} 
  While for the parameters used in this paper the low-lying eigenstates are localized 
  in the thermodynamic limit, in the following we consider systems sizes smaller than the localization length.
\bibitem{not3} 
  The situation is slightly different when a static next-to-nearest neighbor 
  coupling is added (lower right panel).
  In this case, for large values of $\Delta$ and sufficiently large $J_{z0}$, 
  equation~\eqref{eq:teff} does not admit a physically meaningful solution.
\bibitem{note4}
  The raising of ${\delta n}^z_{\pi}$ for $\Delta \gtrsim \bar{\Delta}$ is apparent
  for the case $(I)$, as shown in the upper left panel of Fig.~\ref{fig:deltaN}.
  For the other cases we do not show pertinent data, as this is not the main point we concern with.
\end{thebibliography}
\end{document}